\documentclass[12pt]{article}
\usepackage{amsmath,amssymb,exscale}  
\input epsf

\def\gappeq{\mathrel{\rlap {\raise.5ex\hbox{$>$}}
{\lower.5ex\hbox{$\sim$}}}}
\def\permil{$\%\raise.20ex\hbox{$_0$}}
\def\lappeq{\mathrel{\rlap{\raise.5ex\hbox{$<$}}
{\lower.5ex\hbox{$\sim$}}}}

\begin{document}
\topmargin -1.0cm
\oddsidemargin -0.8cm
\evensidemargin -0.8cm
\pagestyle{empty}
\begin{flushright}
UNIL-IPT-00-08\\
hep-th/0004014\\
April 2000
\end{flushright}
\vspace*{5mm}

\begin{center}

{\Large\bf Localizing Gravity on a String-Like Defect \\
\vskip 0.2cm
in Six Dimensions}\\
\vspace{1.0cm}

{\large Tony Gherghetta\footnote{Email: tony.gherghetta@ipt.unil.ch}
and Mikhail Shaposhnikov\footnote{Email: mikhail.shaposhnikov@ipt.unil.ch}}\\
\vspace{.6cm}
{\it {Institute of Theoretical Physics\\ University of Lausanne\\ 
CH-1015 Lausanne, Switzerland}}
\vspace{.4cm}
\end{center}

\vspace{1cm}
\begin{abstract}

We present a metric solution in six dimensions where gravity
is localized on a four-dimensional singular string-like defect. The 
corrections to four-dimensional gravity from the bulk continuum modes 
are suppressed by ${\cal O}(1/r^3)$. No tuning of the
bulk cosmological constant to the brane tension is required
in order to cancel the four-dimensional cosmological constant.

\end{abstract}

\vfill

\eject
\pagestyle{empty}
%\clearpage\mbox{}\clearpage
\setcounter{page}{1}
\setcounter{footnote}{0}
\pagestyle{plain}

%%%%%%%%%%%%%%%%%%%%%%%%%%%%%%%%

It is an old idea that spacetime may have more than four dimensions,
with extra coordinates being unobservable at available energies. A
first possibility arises in Kaluza-Klein type theories (see
e.g~\cite{KK} and references therein), where  the D-dimensional
metric has the form
\begin{equation}
    ds^2 = g_{\mu\nu}(x^\mu) dx^\mu dx^\nu - \gamma_{ab}(x^a)dx^a dx^b~.
\label{KK}
\end{equation}
Here $g_{\mu\nu}$ is the metric of our four-dimensional world, while
$\gamma_{ab}$ is the metric associated with $D-4$ (small, with a size
$M^{-1}$) compact extra dimensions. The compactness of extra
dimensions makes them unobservable at energies $E<M$, and manifests
itself in the existence of an infinite tower of states with
four-dimensional masses $\sim M$.  

In fact, the Kaluza-Klein metric is not the most general metric
consistent with Poincare invariance in four dimensions.  Its
generalization was proposed in~\cite{RS1}, and is given by 
\begin{equation}
    ds^2 = \sigma(x^a)g_{\mu\nu}(x^\mu)dx^\mu dx^\nu 
         - \gamma_{ab}(x^a)dx^a dx^b~.
\label{conf}
\end{equation}
where $\sigma(x^a)$ is a conformal factor depending on the extra
coordinates only. A number of specific solutions of the Einstein
equations in six-dimensional (6d) spacetime with a positive 6d cosmological
constant were found in~\cite{RS1}, leading to non-compact extra
dimensions while still leaving them unobservable at low energies.

Yet another idea leading to non-compact extra dimensions was
suggested in~\cite{RS2,akama,visser}. In this case the four dimensions of our
world were identified with the internal space of topological defects
residing in a higher-dimensional spacetime (e.g. a domain wall in
5d,  string in 6d, monopole in 7d, instanton in 8d, etc). In these
type  of backgrounds, as a rule, there are fermionic and scalar
zero modes, that can be associated with the four-dimensional particles
that we observe. At that time it was not clear how to  localize the
gauge fields and gravity on topological defects  in order to make the
whole construction realistic.

The solitons of string theory - D-branes - provide a natural
framework for the localization of gauge and matter fields on the 
worldvolume of the branes~\cite{branes}.  In field theory language
the branes can be associated with topological defects.
Moreover, in Ref.~\cite{rusu} it was
discovered that gravity could be localized on 
the 3-brane domain wall in 5d spacetime.
A normalizable graviton zero mode residing on the brane correctly
reproduces 4d gravity,
while the continuum spectrum of 5d gravitons living in the bulk,
gives only a small correction ${\cal O}(1/r^2)$ to Newton's law at
large distances~\cite{rusu}. The metric of the corresponding 5d
spacetime has the general structure of eq. (\ref{conf}).

The aim of the present paper is to see what happens with gravity
around a 3-brane of a specific structure (local string defect in 
field theory language) in 6d spacetime with a negative cosmological 
constant.  In
fact, a regular solution of the Einstein equations in this case for
an empty space follows immediately from~\cite{RS1}, but does not
give any possibility of compactification. However,  the existence of
a brane with positive tension changes the situation  and
we find a solution which is very similar to that of Ref.~\cite{rusu}.
In contrast to 5d case, there is no fine-tuning of the cosmological 
constant in the bulk to the tension of the brane (the origin of this
difference is that the 1d space in the domain wall scenario is flat,
while the 2d space around the string defect can be curved). 
Similarly to the solution in Ref.~\cite{rusu}, there is a normalizible 
graviton zero mode attached to the string-like defect, and the contribution 
of bulk gravitons is suppressed, leading to ${\cal O}(1/r^3)$ 
violations of Newton's law. A hierarchy between the four-dimensional
Planck scale and the Planck scale in 6d can be obtained, 
leading to a solution of the gauge hierarchy problem 
similar to that of ref. \cite{add}.

Other solutions obtained with two transverse dimensions include 
a generalization of  the original 5d domain wall setup
to the case of parallel brane 
sources~\cite{cp}, and the case of global string defects~\cite{ck,rg,ov}.
Furthermore, a class of radially symmetric solutions was 
considered in~\cite{cehs,luty}.

Let us begin with the details of our solution. In 6d the Einstein
equations with a bulk cosmological constant $\Lambda$ and stress-energy
tensor $T_{AB}$ are
\begin{equation}
    R_{AB} - \frac{1}{2} g_{AB} R = \frac{1}{M_6^4}\left(\Lambda 
   g_{AB} + T_{AB}\right)~,
\end{equation}
where $M_6$ is the six-dimensional reduced Planck scale.  We will assume 
that there exists a solution that respects 4d Poincare invarance. 
A six-dimensional metric satisfying this ansatz is
\begin{equation}
\label{metric}
    ds^2 = \sigma(\rho) g_{\mu\nu} dx^\mu dx^\nu -d\rho^2 
    - \gamma(\rho) d\theta^2\, ,
\end{equation}
where  the metric signature is $(+,-,-,-)$. For the two extra spatial
dimensions we have introduced polar coordinates  $(\rho,\theta)$,
where $0\leq \rho < \infty$ and $0\leq \theta < 2\pi$. With our
metric ansatz (\ref{metric}), the general expression for  the 
four-dimensional reduced Planck scale, $M_P$ expressed in terms of $M_6$ is
\begin{equation}
\label{planck}
     M_P^2 = 2\pi M_6^4 \int_0^\infty d\rho\, \sigma\sqrt{\gamma}~.
\end{equation}
The nonzero components of the stress-energy tensor $T^A_B$ are
assumed to  be
\begin{equation}
\label{source}
     T^\mu_\nu = \delta^\mu_\nu f_0(\rho), \quad T_\rho^\rho = 
     f_\rho(\rho), \quad
     {\rm and} \quad T_\theta^\theta = f_\theta(\rho)~,
\end{equation}
where we have introduced three source functions $f_0, f_\rho$ and
$f_\theta$,  which depend only on the radial coordinate $\rho$. Using
the cylindrically  symmetric metric ansatz (\ref{metric}) and the
stress energy tensor (\ref{source}), the Einstein equations become
\begin{eqnarray}
\label{solnset}
    \frac{3}{2} \frac{\sigma^{\prime\prime}} {\sigma} 
    + \frac{3}{4}\frac{\sigma^\prime}{\sigma}\frac{\gamma^\prime}{\gamma}
    -\frac{1}{4}\frac{\gamma^{\prime 2}}{\gamma^2}
    +\frac{1}{2}\frac{\gamma^{\prime\prime}}{\gamma}
    &=& -\frac{1}{M_6^4}(\Lambda + f_0(\rho)) +
    \frac{1}{M_P^2}\frac{\Lambda_{phys}}{\sigma}~,
    \nonumber\\
    \frac{3}{2}\frac{\sigma^{\prime 2}}{\sigma^2}+
    \frac{\sigma^\prime}{\sigma}
    \frac{\gamma^\prime}{\gamma}&=
    & -\frac{1}{M_6^4}(\Lambda + f_\rho(\rho)) 
    +\frac{1}{M_P^2}\frac{2\Lambda_{phys}}{\sigma}~,\nonumber\\
    2\frac{\sigma^{\prime\prime}}{\sigma} + 
    \frac{1}{2}\frac{\sigma^{\prime 2}}{\sigma^2}
    &=& -\frac{1}{M_6^4}(\Lambda + f_\theta(\rho)) +
    \frac{1}{M_P^2}\frac{2\Lambda_{phys}}{\sigma}~,
\end{eqnarray}
where the $^\prime$ denotes differentiation $d/d\rho$. The constant
$\Lambda_{phys}$ represents the physical four-dimensional
cosmological constant, where
\begin{equation}
     R_{\mu\nu}^{(4)} - \frac{1}{2} g_{\mu\nu} R^{(4)} = 
     \frac{1}{M_P^2}\Lambda_{phys} g_{\mu\nu}~.
\end{equation}
By eliminating two of the equations in (\ref{solnset}), the sources
can be related by the equation
\begin{equation}
\label{sourcerel}
    f_\rho^\prime = 2\frac{\sigma^\prime}{\sigma}(f_0-f_\rho) 
      +\frac{1}{2}\frac{\gamma^\prime}{\gamma}(f_\theta-f_\rho)~.
\end{equation}
In the absence of source terms, the discussion of solutions to this
coupled system of differential equations for arbitrary values of
$\Lambda_{phys}$ and $\Lambda >0$ can be found in~\cite{RS1}. 
However, the case $\Lambda <0$ was not considered there 
because the vacuum
solutions lead to noncompact transverse spaces, and therefore using
(\ref{planck}), one cannot obtain a finite value of the Planck scale.
Here we propose adding singular source terms in order to obtain a 
transverse
space with finite volume (which leads to a finite four-dimensional
Planck scale). Thus, the system of equations  (\ref{solnset}) and
(\ref{sourcerel}) describes the generalization of the setup
considered in \cite{RS1}, to the case where source terms are
included. Similar equations of motion in the global string
context, were also considered in
Refs.~\cite{ck,rg,ov}.

Specifically, we will assume that there is a 3-brane at the origin 
$\rho=0$ which is a four-dimensional local string-like topological defect in the
six-dimensional spacetime, and has a nonzero stress-energy tensor
$T^A_B$ parametrized by (\ref{source}). For example, one may think of
the Nielsen-Olesen string solution in the 6d Abelian Higgs model. The
source functions describe a continuous matter distribution within the
core of radius $\epsilon$ and vanish for $\rho>\epsilon$.
At the origin we will require that our
solution satisfies the boundary conditions
\begin{equation}
     \sigma^\prime\big|_{\rho=0} = 0~, \quad (\sqrt{\gamma})^\prime
     \big|_{\rho=0} = 1  \quad {\rm and} 
     \quad \gamma\big|_{\rho=0} = 0~.
\end{equation}
We have set $\sigma(0) =A$, where $A$ is a constant.
Following~\cite{fiu}, we can integrate over the disk of small radius
$\epsilon$ containing the 3-brane, and define various components of 
the string tension per unit length as
\begin{equation}
     \mu_i = \int_0^\epsilon d\rho\, \sigma^2 \sqrt{\gamma}\, 
     f_i(\rho)~.
\end{equation}
where $i = 0,\rho,\theta$. Using the system of equations
(\ref{solnset})  we obtain the following boundary conditions 
\begin{equation}
\label{junc1}
     \sigma\sigma^\prime\sqrt{\gamma} \big|_0^\epsilon = 
       -\frac{1}{2 M_6^4}(\mu_\rho +\mu_\theta)~,
\end{equation}
and
\begin{equation}
\label{junc2}
     \sigma^2(\sqrt{\gamma})^\prime \big|_0^\epsilon = 
       -\frac{1}{M_6^4}(\mu_0+\frac{1}{4}\mu_\rho-\frac{3}{4}\mu_\theta)~,
\end{equation}
where it is understood that the limit $\epsilon\rightarrow 0$ is taken.
By analogy with string defects in four dimensions, $\mu_\rho + \mu_\theta$
can be referred to as the Tolman mass (per unit length)~\cite{christ}.
Its nonzero value in four dimensions gives rise to the Melvin branch 
for local string defects~\cite{verbin}. Similarly, the analogous equation
of (\ref{junc2}) in four dimensions is related to the string 
angular deficit~\cite{verbin}.
Thus, with these general conditions, any metric 
solution to the Einstein equations with 
sources will lead to nontrivial relationships between the components of the string
tension per unit length.

Let us now restrict to the case where the four-dimensional cosmological
constant $\Lambda_{phys} =0$, and look for a solution outside the
core $(\rho > \epsilon)$ of the form
\begin{equation}
\label{solnform}
     \sigma(\rho) = e^{-c \rho}~.
\end{equation}
Note that we have chosen the arbitrary integration constant, which
corresponds to an overall rescaling of the coordinates $x^\mu$,
such that $\displaystyle \lim_{\epsilon\rightarrow 0} \sigma(\epsilon)=1$.
Then, a solution to the coupled set of equations (\ref{solnset}) can be found
with $\gamma(\rho)=R_0^2\sigma(\rho)$ and
\begin{equation}
\label{soln}
     c=\sqrt{\frac{2}{5}\frac{(-\Lambda)}{M_6^4}}~,
\end{equation}
where $R_0$ is an arbitrary length scale that 
can be fixed from eqs. (\ref{junc1}) and (\ref{junc2}).
Clearly, the negative exponential solution (\ref{solnform}) requires
that $\Lambda < 0$. If we now demand that the solution (\ref{soln})
is consistent  with the boundary conditions (\ref{junc1}) and
(\ref{junc2}), the  components of the string tension per unit length
must satisfy
\begin{equation}
\label{sourceeqs}
     \mu_0 = \mu_\theta + A^2 M_6^4~,
\end{equation}
where $\mu_\rho$ remains undetermined. In fact choosing $\mu_\rho=0$ gives
\begin{equation}
\label{tension}
    \mu_\theta = 2 R_0 M_6^4 c~.
\end{equation}
Thus, as long as sources are introduced at the origin $\rho =0$
satisfying (\ref{sourceeqs}),  we obtain a flat Poincare invariant
solution in four dimensions.  Since the solution is already valid for
$\Lambda_{phys}=0$, there is  no need to tune the brane tension to the
bulk cosmological constant,  $\Lambda$ as in the case~\cite{rusu}. 
However, there is still a tuning in order to satisfy (\ref{sourceeqs}).

Having found a solution with a finite volume transverse space
the four-dimensional reduced Planck scale now becomes
\begin{equation}
    M_P^2 =2\pi R_0 M_6^4 \int_0^\infty d\rho\, \sigma^{3/2} 
    = \frac{5\pi}{3} \frac{\mu_\theta}{-\Lambda}M_6^4~,
\end{equation}
where we have used the relation (\ref{tension}). The
inequality $M_6 \ll M_P$ is possible by adjusting the string tension
or the bulk cosmological constant, and thus could lead to a solution
of the gauge hierarchy problem along the lines of \cite{add}.

In order to see that gravity is only localized on the 3-brane, let us
now consider the equations of motion for the linearized metric
fluctuations. We
will only concentrate on the spin-2 modes and neglect the scalar
modes, which needs to be taken into account together with the bending of
the brane~\cite{bend}. The vector modes are massive
as follows from a simple modification of the results in Ref.~\cite{lt}.
For a fluctuation of the form $h_{\mu\nu}(x,z)
= \Phi(z) h_{\mu\nu}(x)$ where $z=(\rho,\theta)$ and $\partial^2 h_{\mu\nu}(x) 
= m_0^2 h_{\mu\nu}(x)$  we can separate the variables by
defining $\Phi(z)= \sum_{lm} \phi_m(\rho) e^{i l \theta}$.  The
radial modes satisfy the equation~\cite{lt}
\begin{equation}
\label{diffop}
     -\frac{1}{\sigma\sqrt{\gamma}}\,\partial_\rho\left[
      \sigma^2\sqrt{\gamma} \,\partial_\rho \phi_m  \right] = 
      m^2 \phi_m~,
\end{equation}
where $m_0^2=m^2 + l^2/R_0^2$ contains the contributions from the 
orbital angular momentum $l$. The differential operator (\ref{diffop}) 
is self-adjoint provided that we impose the boundary conditions
\begin{equation}
\label{bc}
     \phi_m^\prime(0) = \phi_m^\prime(\infty) = 0~,
\end{equation}
where the modes $\phi_m$ satisfy the orthonormal condition
\begin{equation}
       \int_0^\infty d\rho \, \sigma \sqrt{\gamma}\, \phi_m^\ast \phi_n
       = \delta_{mn}~.
\end{equation}
Using the specific solution (\ref{soln}), the differential
operator ({\ref{diffop}) becomes
\begin{equation}
\label{diffeqn}
      \phi_m^{\prime\prime} -\frac{5}{2} c \phi_m^\prime + m^2 e^{c\rho} 
     \phi_m = 0~.
\end{equation}
This equation is the same as that obtained for the
5d domain wall solution~\cite{rusu}, except that the coefficient of
the the first-derivative term is $2$ instead of $5/2$. This 
difference is due to the extra spatial coordinate in the transverse 
space.
When $m=0$ we clearly see that $\phi_0(\rho)$ = constant is a solution.
Since the modes satisfy the orthonormal condition
\begin{equation}
       R_0 \int_0^\infty d\rho \, e^{-\frac{3}{2} c\rho} \phi_m^\ast \phi_n 
       = \delta_{mn}~,
\end{equation}
a wavefunction in flat space can be defined as 
\begin{equation}
   \psi_m = e^{-\frac{3}{4} c\rho} \phi_m~.
\end{equation}
Thus the zero-mode wavefunction becomes
\begin{equation}
      \psi_0(\rho) = \sqrt{\frac{3c}{2R_0}} e^{-\frac{3}{4} c\rho}~,
\end{equation}
which shows that the zero-mode tensor fluctuation is localized near 
the origin $\rho =0$ and is normalizable.

The contribution from the nonzero modes will modify Newton's law on the
3-brane. In order to calculate this contribution we need
to obtain the wavefunction for the nonzero modes at the origin. The 
nonzero mass eigenvalues can be obtained by imposing the boundary conditions
(\ref{bc}) on the solutions of the differential equation (\ref{diffeqn}). 
The  solutions of (\ref{diffeqn}) are
\begin{equation}
    \phi_m(\rho) = e^{\frac{5}{4} c\rho}
    \left[ C_1 J_{5/2}(\frac{2m}{c} e^{\frac{c}{2}\rho}) 
        + C_2 Y_{5/2}(\frac{2m}{c} e^{\frac{c}{2}\rho}) \right]~,
\end{equation}
where $C_1,C_2$ are constants and $J_{5/2}, Y_{5/2}$ are Bessel functions
which can be expressed in terms of elementary functions. 
In the limit that $\rho\rightarrow\infty$, 
the solutions for nonzero $m$ grow exponentially. One way
to regulate this behaviour is to introduce a finite radial distance
cutoff $\rho_{\rm max}$. Then imposing the boundary conditions
(\ref{bc}) at $\rho=\rho_{\rm max}$ (instead of $\rho=\infty$)
will lead to a discrete mass spectrum, where for sufficiently large 
integer $n$ we obtain 
\begin{equation}
      m_n \simeq c(n-\frac{1}{2}) \frac{\pi}{2} e^{-\frac{c}{2} \rho_{\rm max}}~.
\end{equation}
With this discrete mass spectrum we find that
in the limit of vanishing mass $m_n$,
\begin{equation}
      \phi_{m_n}^2(0) = \frac{4}{cR_0} m_n^2 e^{-\frac{c}{2} \rho_{\rm max}}~.
\end{equation}
On the 3-brane the gravitational potential between two point masses
$m_1$ and $m_2$, will receive a contribution from the discrete nonzero modes 
given by
\begin{equation}
      \Delta V(r) \simeq G_N  \frac{m_1 m_2}{r} \sum_n e^{-m_n r}
       \frac{8}{3c^2} m_n^2 e^{-\frac{c}{2} \rho_{\rm max}}~,
\end{equation}
where $G_N$ is Newton's constant.
In the limit that $\rho_{\rm max}\rightarrow \infty$, the spectrum
becomes continuous and the discrete sum is converted into an
integral. Thus the contribution to the gravitational potential becomes
\begin{eqnarray}
       \Delta V(r)
       &\simeq& \frac{16\,G_N}{3\pi c^3} \frac{m_1 m_2}{r}
       \int_0^\infty dm\,m^2 e^{-m r} \\
       &=& \frac{32\,G_N}{3\pi c^3} \frac{m_1 m_2}{r^4}~.
\end{eqnarray}
Thus we see that the correction to Newton's law from the bulk
continuum states grows like $1/r^3$. This correction is more
suppressed than in 5d, because now the  gravitational field of
the bulk continuum modes spreads out in one extra dimension and so
their effect on the 3-brane is weaker.

Some remarks are now in order:

(i) If different components of the brane tension
do not satisfy equation (\ref{sourceeqs}), a more general
solution to the system of equations (\ref{solnset}) can be found along the
lines of ref.~\cite{RS1}. Using the parametrisation $\sigma = z^{4/5}$, and 
$\gamma = \alpha^2\, (z')^2 z^{-6/5}$ (with $\alpha= 4 R_0/(5c)$) 
the general solution can be written as
\begin{equation} 
    z(\rho) = \exp{(-\frac{5}{4} c\rho)} + 2 \beta \sinh
       {(-\frac{5}{4} c\rho)}~,
\end{equation}
where $\beta = 0$ corresponds to the case (\ref{solnform}). The general
condition for the brane tension components now becomes
\begin{equation}
    \mu_0 - \mu_\theta=\beta(\beta+1)(\frac{3}{2}\mu_\theta 
     -\frac{5}{2}\mu_\rho -4\mu_0) 
     +(1+2\beta)^2 A^2 M_6^4~.
\end{equation}
The choice of $\beta <0$ does not lead to any compactification
because $\sigma$ diverges at large $\rho$. However,  $\beta >0$ leads
to non-compact spaces defined for a finite interval
$0<\rho<\frac{2}{5c} \log(\frac{1+\beta}{\beta})$ of the type
discussed in \cite{RS1} that may be used as a description of 
four-dimensional space.

(ii) The metric solution that we have found can also be written in the form
\begin{equation}
    ds^2 = z^2 g_{\mu\nu} dx^\mu dx^\nu - z^2 R_0^2 d\theta^2
         -\frac{4}{c^2 z^2} dz^2~.
\end{equation}
where $z=\exp(-\frac{c}{2} \rho)$.
In this way we see that the origin $\rho = 0$ is now mapped to 
$z=1$. The singular source is spread around the circumference
of a disk of radius $R_0$. This suggests that the 3-brane 
at the origin $\rho=0$ can be interpreted as a wrapped 4-brane 
where all angular points, $\theta$ are identified. In other words,
denoting the wrapped 4-brane by ${\cal M}_4$, then the 3-brane corresponds 
to ${\cal M}_4/S^1$.
While we have given the explicit solution in six dimensions, our
solution can be generalized and 
presumably similar solutions exist at the core of topological
defects in higher dimensions where for $n\geq 2$ transverse dimensions 
the 3-brane can be identified with ${\cal M}_{n+2}/S^{n-1}$, where
${\cal M}_{n+2}$ has $n-1$ coordinates spherically wrapped.
Again, the corrections to 4d gravity on the 3-brane are expected to
be small since the 
bulk continuum modes live in a higher-dimensional space and by Gauss's law
the effects on the 3-brane are suppressed.

(iii) It is also interesting to study whether our solution (or its 
generalization in higher dimensions) can be realized
in an effective supergravity theory. This would be one step towards
embedding the scenario in string theory.

\noindent
{\it Acknowledgments:}
We wish to thank S.~Dubovsky and P.~Tinyakov for 
helpful discussions. This work was supported by the FNRS, 
contract no. 21-55560.98.


\begin{thebibliography}{99}

\bibitem{KK}
{\it Modern Kaluza-Klein Theories}, 
eds. T.~Appelquist, A.~Chodos and P.~G.~Freund,
(Addison-Wesley, 1987).

\bibitem{RS1}
V.~A.~Rubakov and M.~E.~Shaposhnikov,
%``Extra Space-Time Dimensions: Towards A Solution To The Cosmological Constant 
%Problem,''
Phys.\ Lett.\  {\bf B125} (1983) 139.

\bibitem{RS2}
V.~A.~Rubakov and M.~E.~Shaposhnikov,
%``Do We Live Inside A Domain Wall?,''
Phys.\ Lett.\  {\bf B125} (1983) 136.

\bibitem{akama}
K.~Akama, in {\it Proceedings of the Symposium on Gauge Theory and Gravitation}, 
Nara, Japan, eds. K.~Kikkawa, N.~Nakanishi and H.~Nariai (Springer-Verlag, 1983),
[hep-th/0001113].

\bibitem{visser}
M.~Visser,
%``An Exotic Class Of Kaluza-Klein Models,''
Phys.\ Lett.\  {\bf B159} (1985) 22
[hep-th/9910093].

\bibitem{branes}
J.~Polchinski,
%``Dirichlet-Branes and Ramond-Ramond Charges,''
Phys.\ Rev.\ Lett.\  {\bf 75} (1995) 4724
[hep-th/9510017].

\bibitem{rusu}
L.~Randall and R.~Sundrum,
%``An alternative to compactification,''
Phys.\ Rev.\ Lett.\  {\bf 83} (1999) 4690
[hep-th/9906064].

\bibitem{add}
N.~Arkani-Hamed, S.~Dimopoulos and G.~Dvali,
%``The hierarchy problem and new dimensions at a millimeter,''
Phys.\ Lett.\  {\bf B429} (1998) 263
[hep-ph/9803315].

\bibitem{cp}
A.~Chodos and E.~Poppitz,
%``Warp factors and extended sources in two transverse dimensions,''
Phys.\ Lett.\  {\bf B471} (1999) 119
[hep-th/9909199].

\bibitem{ck}
A.~G.~Cohen and D.~B.~Kaplan,
%``Solving the hierarchy problem with noncompact extra dimensions,''
Phys.\ Lett.\  {\bf B470} (1999) 52
[hep-th/9910132].

\bibitem{rg}
R.~Gregory,
%``Nonsingular global string compactifications,''
Phys.\ Rev.\ Lett.\  {\bf 84} (2000) 2564
[hep-th/9911015].

\bibitem{ov}
I.~Olasagasti and A.~Vilenkin,
hep-th/0003300.

\bibitem{cehs}
C.~Csaki, J.~Erlich, T.~J.~Hollowood and Y.~Shirman,
%``Universal aspects of gravity localized on thick branes,''
hep-th/0001033.

\bibitem{luty}
J.~Chen, M.~A.~Luty and E.~Ponton,
%``A critical cosmological constant from millimeter extra dimensions,''
hep-th/0003067.

\bibitem{fiu}
V.~P.~Frolov, W.~Israel and W.~G.~Unruh,
%``Gravitational Fields Of Straight And Circular Cosmic Strings: Relation 
% Between Gravitational Mass, Angular Deficit, And Internal Structure,''
Phys.\ Rev.\  {\bf D39} (1989) 1084.

\bibitem{christ}
M.~Christensen, A.~L.~Larsen and Y.~Verbin,
%``Complete classification of the string-like solutions of the  gravitating Abelian Higgs model,''
Phys.\ Rev.\  {\bf D60} (1999) 125012
[gr-qc/9904049].

\bibitem{verbin}
Y.~Verbin,
%``Cosmic strings in the Abelian Higgs model with conformal coupling to  gravity,''
Phys.\ Rev.\  {\bf D59} (1999) 105015
[hep-th/9809002].

\bibitem{bend}
J.~Garriga and T.~Tanaka,
%``Gravity in the brane-world,''
hep-th/9911055; 
S.~B.~Giddings, E.~Katz and L.~Randall,
%``Linearized gravity in brane backgrounds,''
JHEP {\bf 0003} (2000) 023
[hep-th/0002091].

\bibitem{lt}
G.~V.~Lavrelashvili and P.~G.~Tinyakov,
%``On Possible Spontaneous Compactification Leading To Zero Cosmological Constant. (In Russian),''
Sov.\ J.\ Nucl.\ Phys. {\bf 41} (1985) 172.

\end{thebibliography}
\end{document}